\documentclass[a4paper,fleqn]{cas-dc}

\usepackage[square,numbers,sort&compress]{natbib}
\usepackage[pdftex]{}
\usepackage{epstopdf}
\usepackage[title]{appendix}

\def\tsc#1{\csdef{#1}{\textsc{\lowercase{#1}}\xspace}}
\tsc{WGM}
\tsc{QE}

\begin{document}
\let\WriteBookmarks\relax
\def\floatpagepagefraction{1}
\def\textpagefraction{.001}

\title [mode = title]{Collective scattering in lattice-trapped Sr atoms via dipole-dipole interactions}

\tnotemark[1] 

\author[]{Shengnan Zhang}[]

\ead{s.zhang.3@bham.ac.uk}

\credit{Writing – review \& editing, Writing – original draft, Methodology, Investigation, Formal analysis, Supervision, Project administration}

\affiliation[1]{organization=School of Physics and Astronomy, University of Birmingham,
            addressline={}, 
            city={Edgbaston, Birmingham},
            postcode={B15 2TT}, 
            state={},
            country={United Kingdom}}
\author[]{Sandhya Ganesh}[]
\credit{Writing – review \& editing, Formal analysis}
\author[]{Balsant Shivanand Tiwari}[]
\credit{Writing – review \& editing, Formal analysis}
\author[]{Kai Bongs}[]
\credit{Writing – review \& editing, Writing – original draft, Formal analysis, Methodology}
\author[]{Yeshpal Singh}[]
\credit{Writing – review \& editing, Writing – original draft, Formal analysis, Supervision, Project administration, Methodology, Funding acquisition}

\ead{Y.Singh.1@bham.ac.uk}

\cortext[1]{Corresponding author}
\fntext[1]{}

\begin{abstract}
We investigate, based on the coupled dipole model, collective properties of dense Sr ensembles trapped in a three-dimensional (3D) optical lattice in the presence of dipole-dipole interactions induced on the 5$s5p^{3}$P$_{0}\to5s4d^{3}$D$_{1}$ transition. Our results reveal that the collective scattering properties, such as the scattered light intensity, frequency shift and linewidth, strongly depend on the interatomic distance and the atom number in the lattice. Moreover, the emission intensity is strongly dependent on the atomic distribution in lattices, the laser polarization and the detection position. The results not only offer the understanding of collective behaviors of lattice-trapped ensembles with an atom number equivalent to the experimental scale, but also provide an excellent platform for exploring many-body physics, thereby, opening a new window for applications like quantum information processing and quantum simulation.

\end{abstract}

\makeatletter\def\Hy@Warning#1{}\makeatother

\maketitle
\section{Introduction}\label{i}
Cold and dense atomic gases exhibit collective light scattering due to the presence of strong light-induced dipole-dipole interactions (DDIs) \cite{1,2,3}. Consequently, collective properties, such as collective frequency shift and line broadening, are modified owing to the emesrgence of collective eigenstates of the system \cite{4,5}. DDIs play a crucial role in such a dense ensemble where the interatomic distance is smaller than the photon wavelength, and the real and imaginary components of the interactions respectively account for the modified shift and broadening properties. The study of DDIs is not only beneficial to the understanding of the fundamental physics \cite{6,7}, but also plays a key role in a range of applications, including optical lattice clocks \cite{8,9,10} and topological quantum optics \cite{11,12}.

In past decades, various theoretical models have been established for the DDIs study \cite{13,14,15,16,17}. The random walk (RW) model that considers atoms to be independent accounts for the incoherent scattering, whereas the coupled dipole (CD) model \cite{37} taking coherence into account captures a complete picture of light scattering in a dense sample as well as provides an excellent prediction of experimental results \cite{18,19}. On the experimental side, the unprecedented control on cold atoms enables us to exploit the collective optical response of a dense sample system. For example, collective excitation in a Rydberg blockade ensemble has been observed \cite{20,21,22,23,24,25,26}; Interactions with Rb cold atoms at 780 nm have extensively been reported \cite{27,28,29,30,31}. In addition, a two-dimensional atomic array has recently attracted attention for simulating the cooperative resonance in light scattering \cite{43,44}. However, alkaline-earth metal atoms, in particular, Sr, have not yet attracted much attention in studying dipolar interactions. The long wavelength 2.6 $\mu$m of the 5$s5p^{3}$P$_{0}\to5s4d^{3}$D$_{1}$ transition associated with the short magic wavelength 412.8 nm for a lattice calculated in Ref \cite{32} facilitates DDIs between atoms over several sites of the optical lattice. Moreover, the involved low-lying states make the interaction immune to external fields and other perturbations. These unique strengths of Sr atomic system ensure it to be an excellent candidate for DDIs. However, the current works on Sr are limited to 1D and 2D optical lattices and a low atom number \cite{33,34,35,36}.

In this work, by applying coupled dipole model, we explore the collective scattering arising from strong dipole-dipole interactions induced on 5$s5p^{3}$P$_{0}\to5s4d^{3}$D$_{1}$, with Sr cold atoms trapped in a 3D optical lattice. We study the effect of the interatomic distance and the atom number on scattering properties, such as the frequency shift and linewidth. Our numerical simulations show that collective properties are strongly dependent on the interatomic distance and the atom number. The scattering intensity as a function of the interatomic distance and the laser detuning is compared between different lattices. Furthermore, the dependence of the intensity on the atomic distribution in lattices, the laser polarization and the detection position is investigated.
 
\section{Numerical Treatment}\label{nt}

\begin{figure}
		\includegraphics[width=8.5 cm]{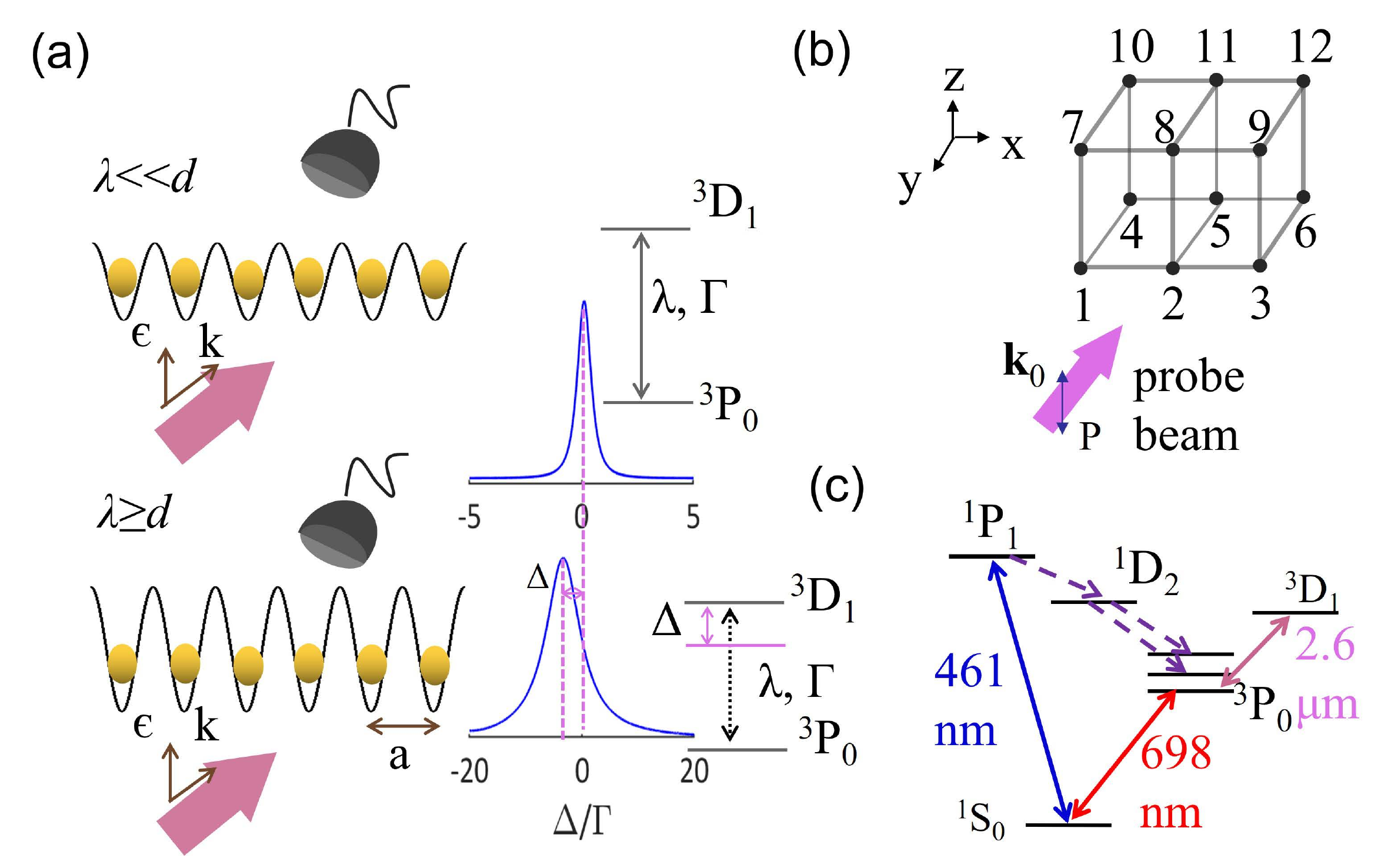}
	  \caption{(a) Schematic of interactions between atoms in an optical lattice in cases of $\lambda\ll d$ and $\lambda\geq d$, where $\lambda$ and $d$ are the transition wavelength and the interatomic distance, respectively, here $d=a$ with $a$ being the lattice spacing. In the case of $\lambda\ll d$, the fluorescence is detected by a detector D when the laser is resonant with the atomic transition $\lambda$; However, the atoms are coupled by DDIs when $\lambda\geq d$, and the fluorescence suffers from linewidth broadening and detuning $\Delta$ relative to the atomic resonance. The atom here is seen as a two-level system with a natural linewidth $\Gamma$. (b) Example of ordered atoms in a $3\times2\times2$ optical lattice. The polarization P and wavevector $k_{0}$ of the probe beam are along $\hat{z}$ and $\hat{y}$ directions, respectively. (c) Simplified relevant Sr energy levels.}\label{fig1}
\end{figure}

In a lattice-trapped sample, atoms undergo various interactions which depend on the spacing a relative to the wavelength $\lambda$ of the driving laser. In dense gases where atoms are sufficiently closely spaced, i.e., $\lambda\geq d$, dipole-dipole interactions between atoms become pronounced giving rise to a level shift and line broadening. However, in dilute gases, where $\lambda\ll d$, collective effects are negligible due to weak dipolar interactions, instead collisional interactions dominate. The comparison between the two situations is depicted in Fig.~\ref{fig1}. The laser propagating vertically towards the lattice-trapped atoms under $\lambda\ll d$ and $\lambda\geq d$ respectively drives interactions between the atoms. The detected fluorescence spectra show different optical properties due to different interaction mechanisms.

Here, we utilize the coupled dipole model incorporating position dependent correlations between atoms \cite{19} and multiple scattering events \cite{30,37} to capture a complete picture of coherent effects of a dense Sr ensemble. We focus on an ensemble trapped in a 3D optical lattice in a Mott insulator where each atom can be seen as a two-level system by applying a static magnetic field to lift the degeneracy of the Zeeman manifolds to induce interactions only on $|g\rangle=^{3}$P$_{0}$, $m=0\to|e\rangle$=$^{3}$D$_{1}$, $m\prime=0$. Initially the atoms are assumed to be in the ground state $|g\rangle$. A weak-intensity probe laser with frequency $\omega$ illuminates dipoles (with the resonance frequency $\omega_{0}=2\pi/\lambda$) localized in the lattice. Driven by the incident laser, each dipole is coupled to all others due to DDIs and emits photons as an emitter.

In the coupled dipole model, $\hat{z}$-polarized probe laser is propagating along $\hat{y}$ direction. The dynamics of an open quantum system consisting of $N$ atoms is governed by the master equation for the reduced atomic density matrix $\rho$ is
\begin{equation}
\centering
\dot{\rho}=-\frac{i}{h}[H,\rho]+D(\rho)\label{eq1}
\end{equation}
where the Hamiltonian $H$ takes into account the coupling of dipoles and the radiation field, characterized by \cite{13} 
\begin{equation}
\begin{aligned}
H=\sum_{i,\alpha}\Delta^{\alpha}b_{i}^{\alpha^{\dagger}}b_{i}^{\alpha}+\sum_{i,\alpha}\Omega^{\alpha}(e^{ik_{0}r_{i}}b_{i}^{\alpha^{\dagger}}+e^{-ik_{0}r_{i}}b_{i}^{\alpha})\\
+\sum_{i\neq j,\alpha,\alpha'}V_{i,j}^{\alpha,\alpha'}b_{i}^{\alpha^{\dagger}}b_{i}^{\alpha'}\label{eq2}
\end{aligned}
\end{equation}
where $\Delta^{\alpha}$ denotes the detuning between the transition resonance and the laser with the polarization $\alpha$, wave vector $k_{0}$ and Rabi frequency $\Omega^{\alpha}$ given by $\Omega^{\alpha}\equiv\zeta E_{0}/\hbar$ with $\zeta$ being the transition dipole moment and $E_{0}$ the amplitude of the laser; $r_{i}$ denotes the position vector of the $i^{th}$ dipole; $b_{i}^{\alpha}$ is an atomic transition operator of the transition to |$\alpha\rangle$ for the $i^{th}$ dipole defined by $b_{i}^{\alpha}\equiv|\alpha_{i}\rangle\langle g_{i}|$, here |$g\rangle$ and |$\alpha\rangle$ depict $^{3}$P$_{0}$ and $^{3}$D$_{1}$ states, respectively; $V_{i,j}^{\alpha,\alpha'}$ describes the dispersive interactions between the atom $i$ and $j$.

The second term $D(\rho)$ in Eq.~(\ref{eq1}) is described by \cite{37}
\begin{equation}
\begin{aligned}
D(\rho)=\sum_{i,j,\alpha,\alpha'}\Gamma_{i,j}^{\alpha,\alpha'}(2b_{j}^{\alpha'}\rho b_{i}^{\alpha^{\dagger}}-\{b_{i}^{\alpha^{\dagger}}b_{j}^{\alpha'},\rho\})\label{eq3}
\end{aligned}
\end{equation}
where $\Gamma_{i,j}^{\alpha,\alpha'}$ is responsible for the dissipative interaction and accounts for the superradiant emission. 

The dispersive and dissipative interactions, respectively, $V_{i,j}^{\alpha,\alpha'}$ and $\Gamma_{i,j}^{\alpha,\alpha'}$, are closely connected and compose the real and imaginary parts of dipolar interactions $G_{i,j}^{\alpha,\alpha'}$ between the atom $i$ and $j$, respectively, separated by $r_{i,j}$. The former term leads to the frequency shift and the latter gives rise to the linewidth broadening of the fluorescence spectrum. The dipolar interaction is given by $G_{i,j}^{\alpha,\alpha'}$, seen in Appendix~\ref{AppA}. In order to calculate the value of $\hat{r}_{i,j}^{\alpha}$, we set up a coordinate system shown in Fig.~\ref{fig1}(b).

In the weak excitation limit ($\Omega^{\alpha}\ll\Gamma$), the atomic coherence $b_{i}^{\alpha}$ can be deduced by solving the master equation in the steady state ($\frac{db_{i}^{\alpha}}{dt}=0$). To an approximation, the condition of $\rho_{G,G}=$Tr$[\hat{\rho}|G\rangle\langle G|]=1$, $\rho_{j\alpha,m\beta}=$Tr$[\hat{\rho}(\hat{b}_{m}^{\beta\dagger}\hat{b}_{j}^{\alpha})]=0$ is satisfied and $b_{j}^{\alpha}\equiv\rho_{j\alpha,G}=$Tr$[\hat{b}_{j}^{\alpha}\hat{\rho}]$. By ignoring the multiatomic coherence, we obtain the atomic coherence $b_{i}^{\alpha}$ of the atom $i$ to all others along $\alpha$, expressed by \cite{38,39,40,41,42}
\begin{equation}
\begin{aligned}
b_{i}^{\alpha}=\frac{\Omega^{\alpha}\delta_{\alpha,\beta}e^{i\mathbf{k_{0}\cdot r_{i}}}/2}{\Delta^{\alpha}+i\Gamma/2}+\sum_{j\neq i,\alpha'}\frac{G_{i,j}^{\alpha,\alpha'}}{\Delta^{\alpha}+i\Gamma/2}b_{j}^{\alpha'}\label{eq6}
\end{aligned}
\end{equation} 
where the probe laser is polarized along $\beta$, the $i^{th}$ atom is located at the position $\mathbf{r_{i}}$.

By summing the product of the atomic coherence $b_{i}^{\alpha'}b_{j}^{\alpha^{*}}$ over the entire space, the fluorescence intensity measured at the position $r_{s}$ in the far field is obtained by \cite{42}
\begin{equation}
\begin{aligned}
I(r_{s})=\frac{9\hbar^{2}\Gamma^{2}}{16k^{2}\mu^{2}r_{s}^{2}}\sum_{i,j}e^{-i\mathbf{k_{s}}\cdot \mathbf{r_{i,j}}}\sum_{\alpha,\alpha'}(\delta_{\alpha,\alpha'}-\mathbf{\hat{r}_{s}^{\alpha}\hat{r}_{s}^{\alpha'}})b_{i}^{\alpha'}b_{j}^{\alpha^{*}}\label{eq7}
\end{aligned}
\end{equation} 
where $\mu$ is the atomic transition dipole moment; $\mathbf{r_{s}}$ is the positions of the detection; $b_{j}^{\alpha^{*}}$ is the complex conjugate of $b_{j}^{\alpha}$.

Therefore, the model enables us to calculate the fluorescence intensity emitted from lattice-trapped Sr ensembles for various interatomic distances and atom numbers based on Eqs.~(\ref{eq6}) and (\ref{eq7}). The frequency shift and the linewidth can be deduced by measuring the resonant frequency and the full width at half maximum (FWHM) of the fluorescent spectra, respectively.

\section{Results and Discussion}\label{rd}
In this work we aim to investigate scattering properties, such as the intensity, the frequency shift and the linewidth, from a dense Sr ensemble in a 3D optical lattice due to dipole-dipole interactions on $^{3}$P$_{0}\to^{3}$D$_{1}$. The branching ratio of $^{3}$D$_{1}\to^{3}$P$_{0}$ is 60$\%$, which is dominant compared to two other transitions $^{3}$D$_{1}\to^{3}$P$_{1,2}$. We compare parameters of Sr ensembles with the extensively studied Rb atomic ensemble listed in Tab.~\ref{tbl1}. We studied the effect of the interatomic distance and the total atom number on the scattering. We also exploited the dependence of the intensity on the atomic distribution in lattices, the laser polarization and the detection position. Intensities as a function of $d$ and $\Delta$ are also calculated based on Eq.~(\ref{eq7}).

\begin{table}
\caption{Comparison of parameters between a Rb ensemble and a Sr ensemble for dipole-dipole interactions. The value of $d$ for Sr $^{1}$S$_{0}\to^{1}$P$_{1}$ and $^{1}$S$_{0}\to^{3}$P$_{1}$ are calculated from Ref \cite{42}. Other values are 
cited from Refs \cite{18,21}. In this work, Sr ensembles are assumed to be trapped in a Mott insulator with a lattice spacing of 206 nm.
}\label{tbl1}
\begin{tabular*}{\tblwidth}{@{}LLLLL@{}}
\toprule
\toprule
  &Transitions&$\lambda$ (nm)&$d$ (nm)&$\lambda/d$  \\
\midrule
 &$^{3}$P$_{0}\to^{3}$D$_{1}$&$2600$&$206$ &$12.6$ \\
 Sr&$^{1}$S$_{0}\to^{1}$P$_{1}$ &$461 $&$1800$ \cite{42}&$0.26$\\
 & $^{1}$S$_{0}\to^{3}$P$_{1}$&$689$&$1800$ \cite{42}&$0.38$\\
 \midrule
 &5S$_{1/2}\to5$P$_{3/2}$&$780$&$156$ \cite{18} &$5$ \\ Rb&5S$_{1/2}\to5$P$_{1/2}\to58$D$_{3/2}$&$795$ and $474$ &$3600$ \cite{21}&$0.13$\\
\bottomrule
\bottomrule
\end{tabular*}
\end{table}

\begin{figure}
	\centering
	\includegraphics[width=7.6 cm]{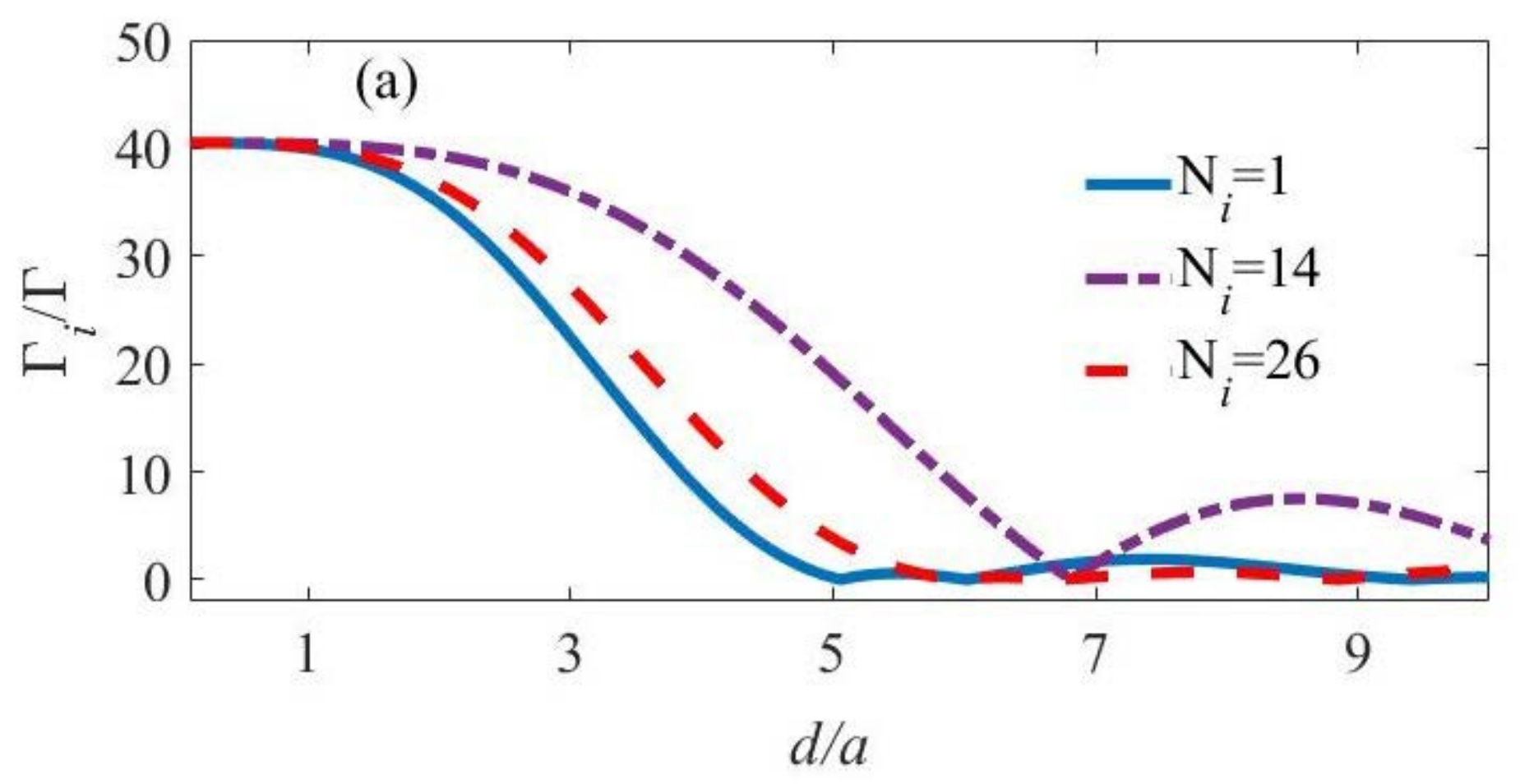}
 \includegraphics[width=8.4 cm]{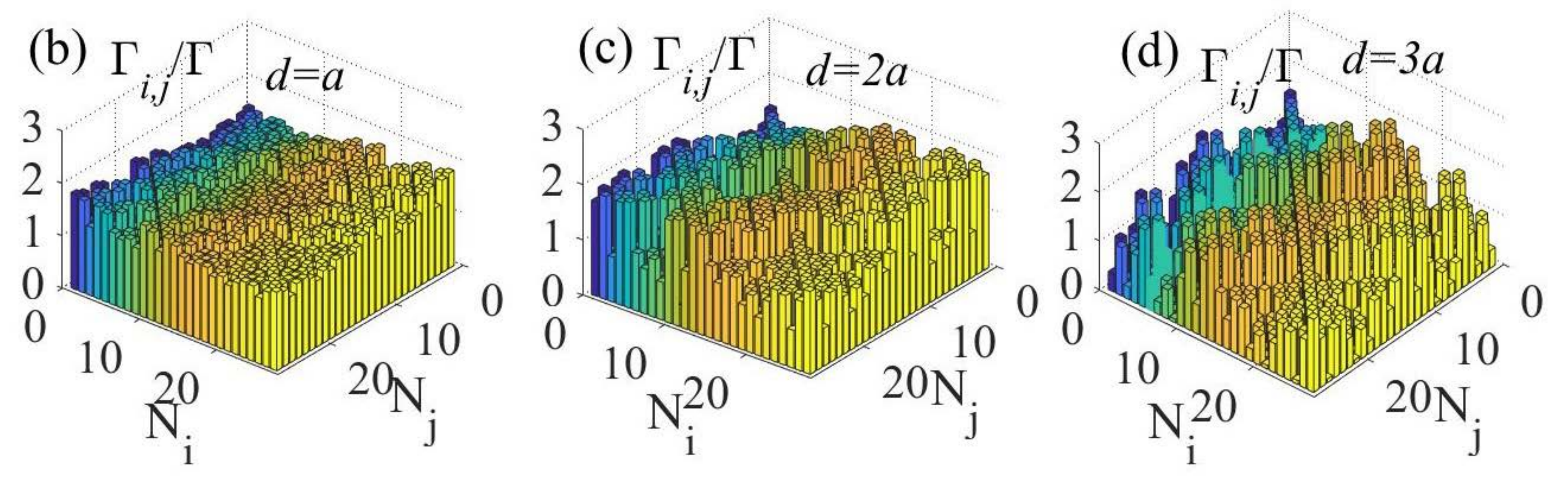}
	  \caption{Damping rate $\Gamma_{i,j}$ between the atoms $i$ and $j$ when $d$ is varied for a $3\times3\times3$ cubic optical lattice. (a) The damping rate $\Gamma_{i}$ of the $i^{th}$ ($i=1, 14, 26$) atom which is the sum of $\Gamma_{i,j}$ between the $i^{th}$ atom and the $j^{th}$ ($j\neq i$) atom as a function of $d$, shown by blue solid, purple dash-dotted, red dash curves, respectively. (b)-(d) show $\Gamma_{i,j}$ when $d=a, 2a, 3a$, respectively. The amplitudes of $\Gamma_{i}$ and $\Gamma_{i,j}$ both are normalized to $\Gamma$ being a single atom’s natural linewidth.}\label{fig2}
\end{figure}

Starting with a $3\times3\times3$ optical lattice, we study the damping rate $\Gamma_{i,j}$ between atoms $i$ and $j$. We assume each site is occupied by a single atom, i.e., Mott insulator states, with the interatomic distance $d$ between any two nearest atoms. We denote the atom number by $N$, which throughout the work is equal to the number of lattice sites as per this single occupation condition. All atoms are coupled to each other with a damping rate $\Gamma_{i,j}$ (see Appendix~\ref{AppA}) under the effect of probing laser $k_{0}$. By solving Eq.~(\ref{eq9}), we calculate $\Gamma_{i,j}$ for different values of $d$ depicted in Fig.~\ref{fig2}. To see the different damping rates of atoms positioned at different sites in the lattice, we compare $\Gamma_{i}$ of the 1$^{st}$, 14$^{th}$ and 26$^{th}$ atom shown in Fig.~\ref{fig2} (a). Here, $\Gamma_{i}=\sum_{j\neq i}\Gamma_{i,j}$, which incorporates the gross contribution from all other atoms. It shows $\Gamma_{i}$ decreases as $d$ increases for all three atoms, and then damply oscillates as $d$ continues to increase. The first atom is located at the corner of the cubic lattice which has only three nearest-neighboring atoms, giving rise to reduced $\Gamma_{i}$. However, the 14$^{th}$ atom, in the center of the lattice, experiences a stronger coupling from nearest-neighboring atoms such that $\Gamma_{i}$ is enhanced. The position of the 26$^{th}$ atom is at the side of the lattice which leads to the value of $\Gamma_{i}$ being the value in between. Fig.~\ref{fig2} (b)-(d) depict $\Gamma_{i,j}$ between the atom $i$ and $j$ ($i,j=1,\cdots,27$) for $d=a, 2a$ and $3a$, respectively. For $\Gamma_{i,i}$, which is shown by the diagonal in each figure, it’s easy to prove that $\Gamma_{i,i} = 3\Gamma/2$.

The physics of DDIs reveals that the excited state $^{3}$D$_{1}$ in our case exhibits a line shift relative to the non-interacting situation when the atoms are spaced closely due to the emergence of strong dipolar interactions \cite{37}. Moreover, the closer are atoms to each other, the larger is the frequency shift as a result of enhanced interactions. For a fixed $N$, as $d$ decreases introducing higher density, the dipolar interaction becomes stronger modifying collective properties of the scattered light.

\begin{figure}
	\centering
 \includegraphics[width=8.5 cm]{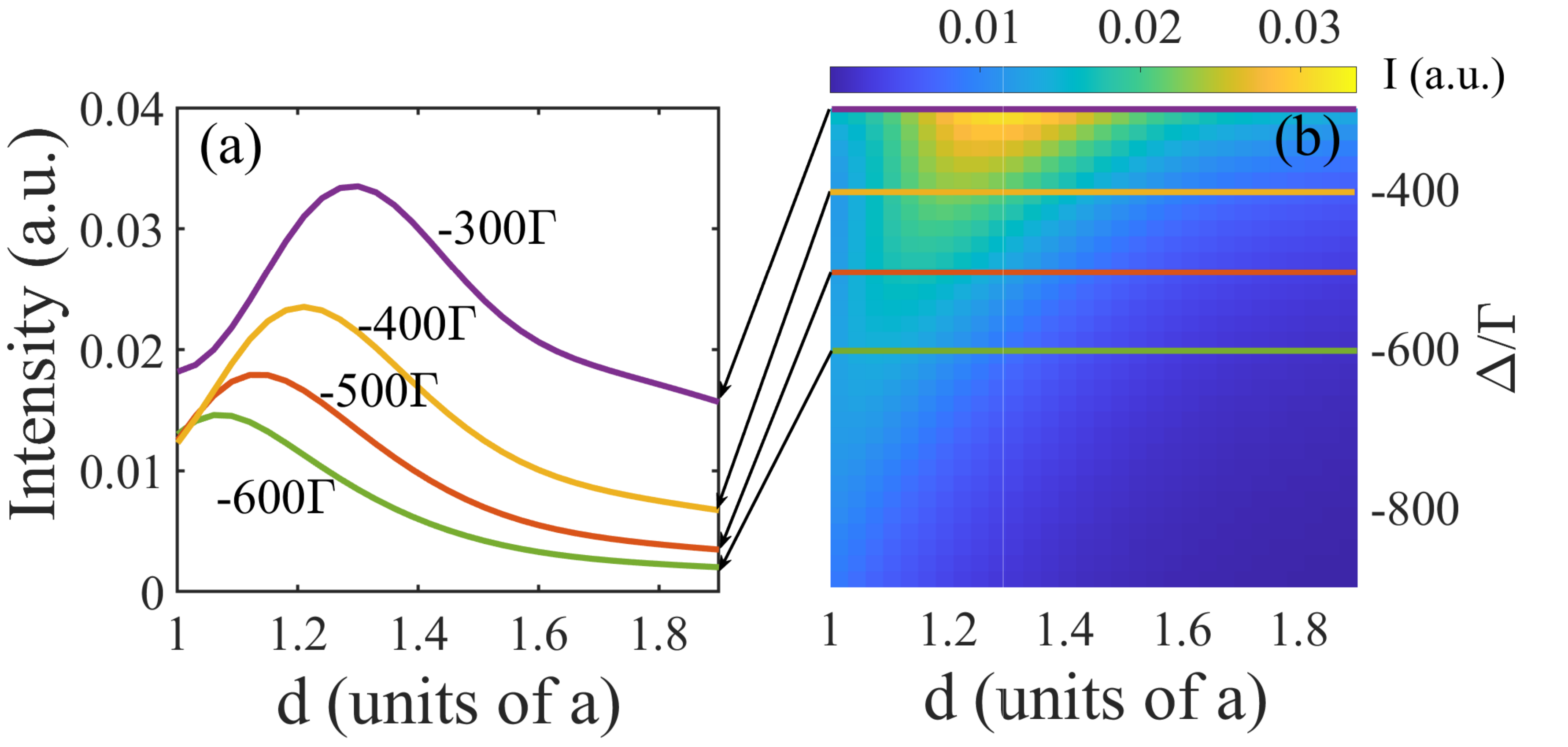}
	  \caption{(a) Scattering intensity as a function of the interatomic distance $d$ at $\Delta=-300\Gamma, -400\Gamma, -500\Gamma, -600\Gamma$ in a $11\times11\times11$ lattice. The laser with a larger detuning is resonant with atoms at a smaller interatomic distance due to the larger energy shift at smaller $d$. The peak intensity at smaller detunings is higher. (b) The chromatogram indicates the scattering intensity as a function of $\Delta/\Gamma$ and $d$. The yellow colour represents a higher intensity while blue lower intensity.}\label{fig3}
\end{figure}

Fig.~\ref{fig3} shows the dependence of the scattering intensity on the interatomic distance $d$ for different laser detunings for a $11\times11\times11$ optical lattice. When the state is shifted, the peak intensity is not at the atomic resonance. One can see that, in Fig.~\ref{fig3} (a), by changing the laser detuning from $-300\Gamma$ to $-600\Gamma$, the peak of the intensity is moving towards a smaller value of $d$. In addition, the amplitude of the scattered intensity at smaller $d$ (equivalent to the larger frequency shift) is reduced due to the stronger interaction between atoms. It’s worth pointing out that the spectra width narrowed down at decreased $d$ is accounted for by the larger frequency-shift gradient while atoms are localized closer to each other. Fig.~\ref{fig3} (b) shows the scattering intensity as a function of $d$ and the laser detuning $\Delta$. The four straight lines indicate the intensity versus $d$ at $\Delta=-300\Gamma, -400\Gamma, -500\Gamma, -600\Gamma$, which correspond to the curves in Fig.~\ref{fig3} (a). It can also be seen that the peak intensity takes place at the point of ($1.3a, -300\Gamma$).

\begin{figure}
	\centering
 \includegraphics[width=7 cm]{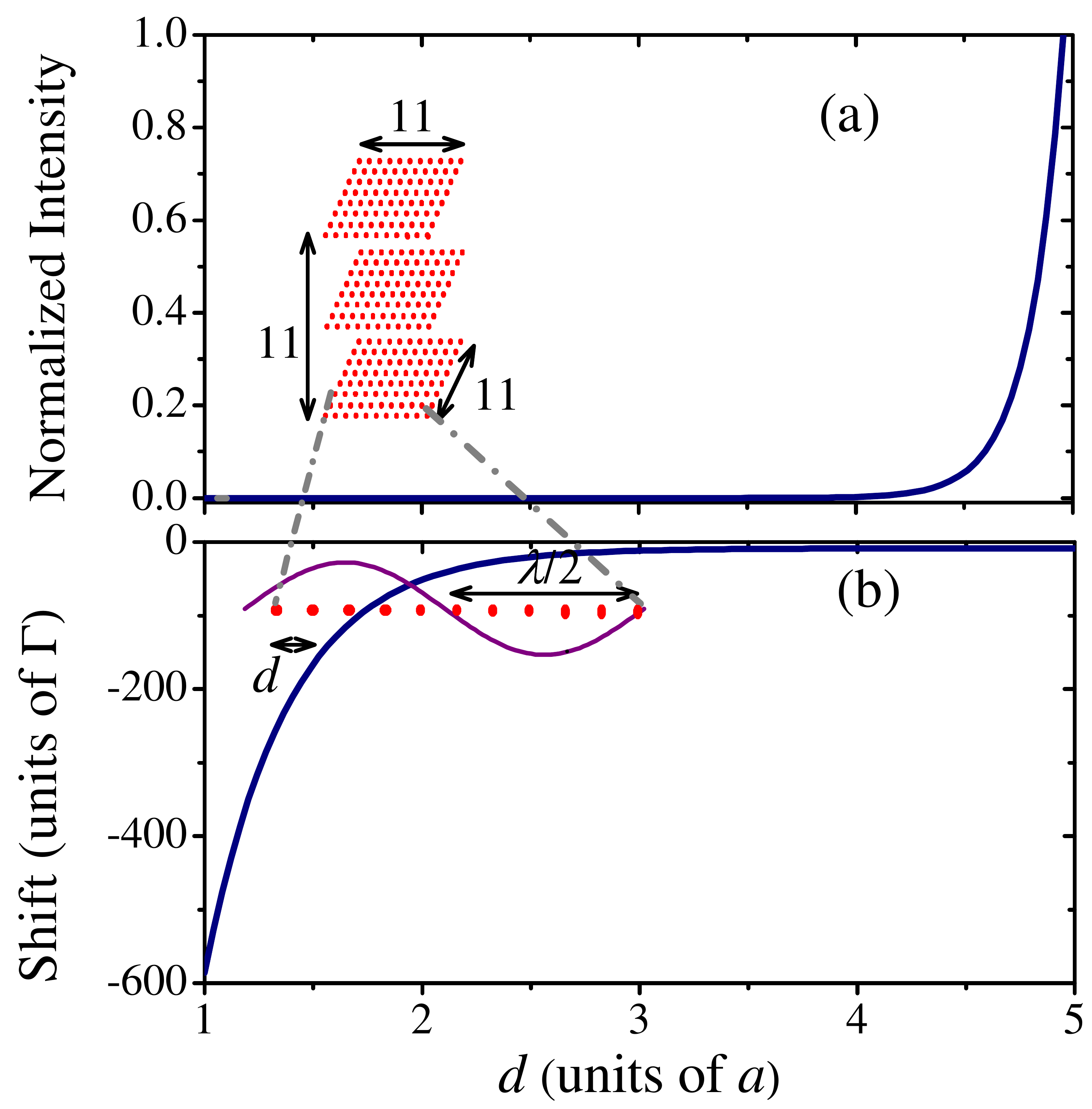}
	  \caption{(a) Scattering intensity as a function of $d$ in a range of 1$\sim 1.7a$ for a $11\times11\times11$ optical lattice. The intensity increases as $d$ increases. (b) Frequency shift versus $d$. The inset is a schematic of $11\times11\times11$ optical lattice with a comparison of the interatomic distance to the laser wavelength. The detection position is along $\hat{z}$ direction.}\label{fig7}
\end{figure}

In addition, we investigate the dependence of the resonant scattering including the intensity and the frequency shift on the interatomic distance $d$ for a $11\times11\times11$ lattice shown in Fig.~\ref{fig7}. By simulating the fluorescent spectra based on Eq~(\ref{eq7}), one can obtain the peak intensity and the frequency shift for $d=a$ to 1.7$a$. The dependence of DDIs on the interatomic distance $d$ has been confirmed by the result shown in Fig.~\ref{fig7}, that the frequency shift decreases when increasing $d$. Fig.~\ref{fig7} (a) demonstrates that the intensity starts to sharply increase at $d=1.4a$, which dramatically differs for various atom numbers. Here the resonant scattering refers to the the light scattered by dipolar interactions at a resonant frequency. The schematic of the lattice is illustrated in the inset. The frequency shift reduces and gradually approach the non-interacting case shown in Fig.~\ref{fig7} (b). At $d=a$, the frequency shift can be as high as $-600\Gamma$, i.e., $\sim$120 MHz.

\begin{figure}
	\centering
 \includegraphics[width=7.1 cm]{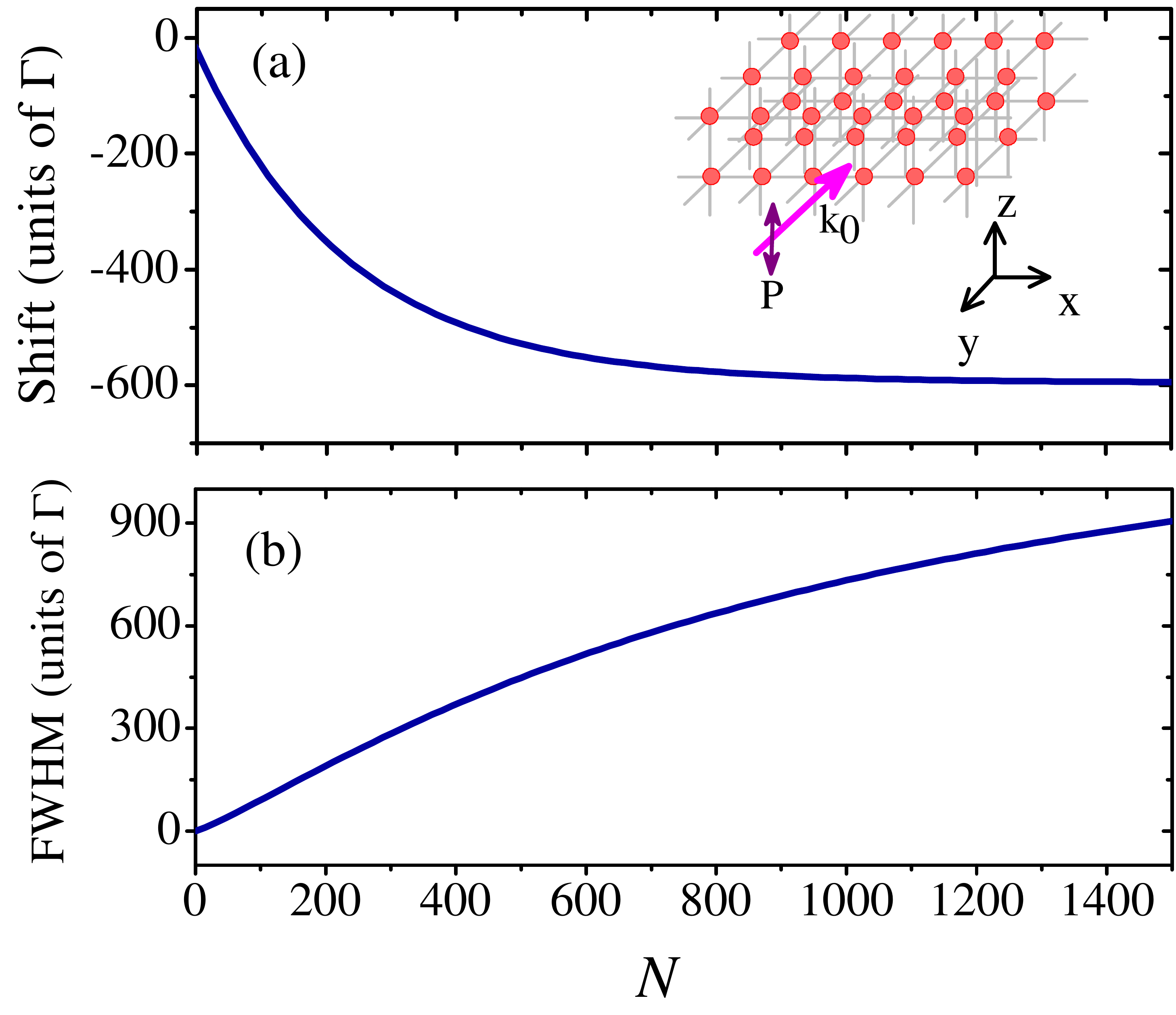}
	  \caption{(a) Resonant frequency shift as a function of $N$ up to 1331 when $d=a$ via dipolar interactions in a cubic lattice. (b) The FWHM of the scattering versus $N$. The solid triangle represents the simulation result and the curve means the fit to data. The insets show the fit to simulation results when $N<36$. The schematic of an optical lattice is shown in the inset.
}\label{fig4}
\end{figure}

Fig.~\ref{fig4} demonstrates the scattering as a function of the atom number $N$ when $d=a$. The lattice in the simulation is cubic with an equal atom number in each direction. The resonant frequency is red-shifted with the increase of $N$ shown in Fig.~\ref{fig4} (a). When $N\geq1000$, the frequency shift is increasing stably and finally approaching $-600\Gamma$. To see the more precise limit value of the frequency shift in a cubic lattice, a larger atom number is needed in the simulation. Meanwhile, the scattering linewidth is broadened shown in Fig.~\ref{fig4} (b). It can be broadened to be $800\Gamma$ at $N=1400$. The results reveal that the interaction varies significantly with $N$ at a small number, e.g., $N<800$, while the variation is slowed down when $N\geq800$. The fit for $N<40$ can be seen in insets. The results reflect the collective scattering features for the further experimental investigation in optical lattices.  

\begin{figure}
	\centering
 \includegraphics[width=17 cm]{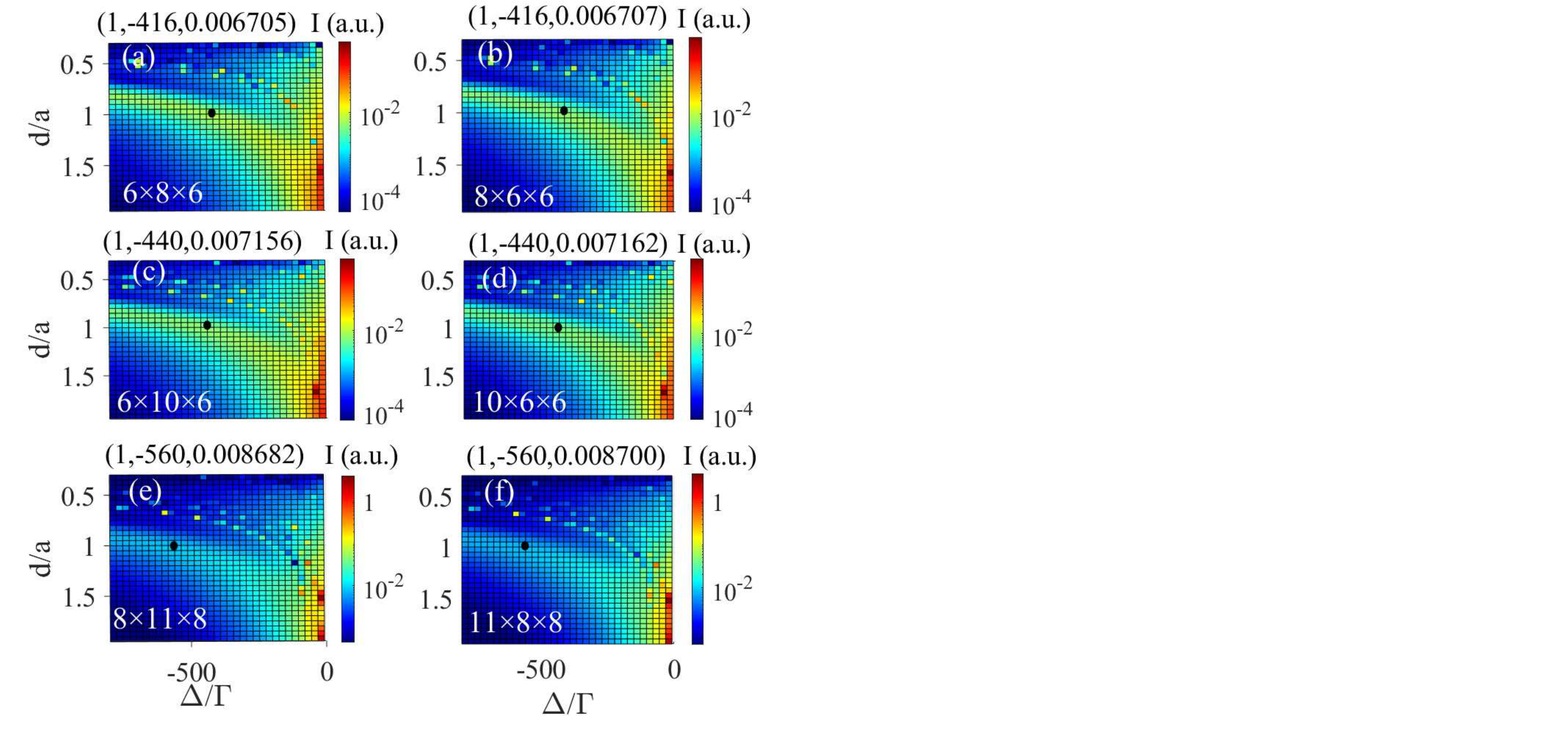}
	  \caption{Plot of the scattered light intensity as a function of the interatomic distance $d$ and laser detuning $\Delta$ for different lattices. The left-bent branches show red-detuned superradiant peaks. The bright thin dotted curves represent subradiant peaks. The intensity at resonance indicates the collisional interaction between atoms. The point of the resonant interaction at $d=a$ is labelled by a black dot and the number for each point is shown on the top of each. The detailed fluorescence spectra of the scattering at $d=a$ are shown in Fig.~\ref{fig6}.}\label{fig5}
\end{figure}

Furthermore, we have numerically calculated the scattering intensity as a function of the interatomic distance $d$ and the laser frequency detuning $\Delta$ for different optical lattices including $6\times8\times6$, $8\times6\times6$, $6\times10\times6$, $10\times6\times6$, $8\times11\times8$ and $11\times8\times8$. The result is shown in Fig.~\ref{fig5}. The fact that the dipole-dipole interaction is dependent on both the atom number and the atomic distribution has been demonstrated by the left-bent branch in each sub-figure. By the comparison of (a) and (b), or (c) and (d), or (e) and (f), the effect of the atomic distribution in a lattice on DDIs can be understood by the difference of scattering intensities in each pair of lattices. The points of resonant interactions at $d=a$ are marked by black dots and the numbers are shown on the top of each sub-figure. It reveals that the scattering is stronger in the direction of more atom number which is different from the directions of the laser polarization and the propagation. The most bright around the zero detuning shows the non-interacting scattering which is stronger than the collective scattering. The dependence of DDIs on the atom number is clear to see by the intensity difference by comparison of, for example, Fig.~\ref{fig5} (a) and (e), and a larger resonance frequency shift in Fig.~\ref{fig5} (e). In addition, a series of narrow subradiant peaks for each lattice can be seen by thin dotted curves which are due to the fact that the dispersive and dissipative interactions do not share the same set of eigenstates \cite{32}.

\begin{figure}
	\centering
 \includegraphics[width=8.2 cm]{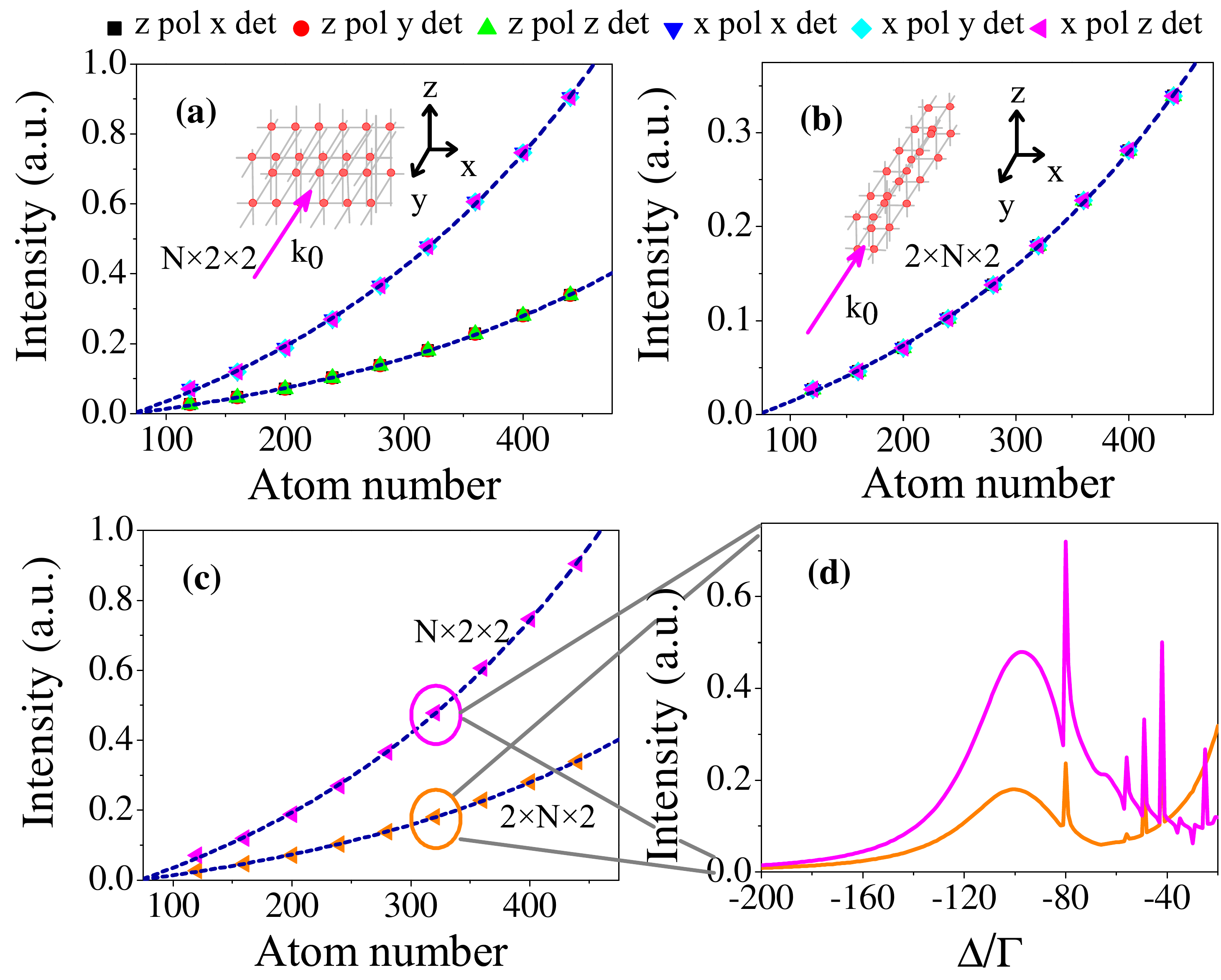}
	  \caption{Dependence of the scattering light intensity on the laser polarization, the atomic distribution in lattices and the detection position as a function of atom number $N$. (a) The scattering intensity as a function of the atom number for the lattice $N\times2\times2$. The laser wave vector is along $\hat{y}$. (b) The scattering intensity as a function of the atom number for the lattice $2\times N\times2$. (c) The comparison of the scattering intensity when $\hat{x}$-polarized $\hat{z}$-detection for the two lattices. (d) The fluorescence spectra when $N=80$ for two lattices. Sharp peaks superpositioned in a wide spectrum indicate the subradiance. The wide spectrum indicates the superradiance.}\label{fig8}
\end{figure}

Finally, we study the dependence of scattering intensity on the laser polarization, the atomic distribution in a lattice and the detection position. The atom number is varied only in one direction while fixed to 2 in each of other two directions. The interatomic distance $d$ is fixed at $a$. According to Eqs.~(\ref{eq6}) and ~(\ref{eq7}), we know that the atomic coherence and the scattering intensity depend on the the laser polarization $\beta$. Here we compare two different laser polarizations (along $\hat{x}$ and $\hat{z}$), two different atomic distributions (along $\hat{x}$ and $\hat{y}$) and three different detection positions in $\hat{x}$, $\hat{y}$ and $\hat{z}$. The result is shown in Fig.~\ref{fig8}. Fig.~\ref{fig8} (a) shows that the scattering intensity is significantly improved for the polarization in the same direction as where the lattice has more trapped atoms, which can be explained by the more coherence when they are along the same direction. Moreover, for the $\hat{x}$-polarization, the intensity increases with the atom number more sharply. The detection position has little influence on the scattering intensity. However, the conclusion above doesn't hold for the case when the direction of the laser wave vector has more atoms shown in Fig.~\ref{fig8} (b). The scattering intensity is equal whatever the detection position and the polarization are, which has the same intensity as the case of the $\hat{z}$-polarization in Fig.~\ref{fig8} (a). In Fig.~\ref{fig8} (c), the intensity in the case of $\hat{x}$-polarization $\hat{z}$-detection for two different configurations of lattices is compared, which shows the lattice with more atoms in $\hat{x}$ scatters stronger light than in $\hat{y}$. The spectra of the scattering for lattices $80\times2\times2$ and $2\times80\times2$ are shown in Fig.~\ref{fig8} (d).

As a direct result of the many-body system governed by the master equation (1) consisting of coherent and dissipative aspects, we observe the emergence of superradiance and subradiance (Fig.~\ref{fig5}$\&$\ref{fig8}). Superradiance arises out of the resonant nature of the two level ($^{3}$P$_{0}\to^{3}$D$_{1}$) atomic ensemble. It is not surprising that the power spectrum (Fig.~\ref{fig8}) of emission behaves like $N^{2}$, $N$ being atom number. Emission line shift and the broadening are direct signatures of strong interaction between two atoms. An excitation created can propagate over several lattice sites. Riding the superradiant structures, there are subradiance peaks (Fig.~\ref{fig5}$\&$\ref{fig8}(d)).  These peaks occur as a consequence of the fact that the dispersive (coherent) and dissipative parts of the master equation~(\ref{eq1}) do not share the same set of eigen states. It allows admixing of symmetric states with collective states, thereby making them excitable by the dipole inducing 2.6 micron laser. It is amply clear that ultra cold neutral alkaline earth atoms like Sr provide a unique platform for studying long ranged many-body physics \cite{52,53,54}. This allows precise control and manipulation on a subwavelength scale. These systems are also excellent for various novel applications such as quantum computing, quantum information due to the strength like immune to the environmental field by the use of low-lying states. As a result of the enhanced cooperative radiation, the superradiance has promising applications in ultranarrow-linewidth superradiant lasing \cite{45,49,50,51}, sensitive gravimeters \cite{46} and quantum information \cite{47}. Its counterpart, the long-lived subradiance, is perfect in quantum memory \cite{48} by taking advantage of its inhibited collective radiation characteristic.

\section{Conclusion}\label{c}
In conclusion, we have performed a detailed theoretical simulation, based on the coupled dipole model, for the optical response of dense Sr atoms in a 3D optical lattice in the presence of dipole-dipole interactions on the transition 5$s5p^{3}$P$_{0}\to5s4d^{3}$D$_{1}$. We focused on the fluorescent spectral characteristics, mainly the scattering intensity, the frequency shift and the linewidth broadening. The numerical analysis shows that the interatomic distance and the atom number have strong impact on the collective scattering in Sr atomic system. In a cubic lattice-trapped Sr ensemble, the frequency shift and the linewidth broadening of the transition can reach hundreds of the natural linewidth, which shows a powerful evidence of the observation of dipolar interactions. Moreover, the dependence of the scattering intensity on the atomic distribution in optical lattices, the laser polarization and the detection position is investigated. The intensity is significantly enhanced for the polarization parallel to the direction of more atoms of the lattice which differs from the wave vector direction. This work opens a window to the insight into the collective behavior of a lattice-trapped Sr ensemble with an experiment equivalent atom number due to dipolar interactions.

\printcredits

\section*{Acknowledgements}
We thank Dr Dimitri M Gangardt and Dr Samuel Lellouch for the fruitful discussion and helpful proof reading. SG and BT acknowledge funding from European Union's Horizon 2020 Research and Innovation Programme - Grant Agreement No. 860579 (MoSaiQC Project).

\begin{appendices}

\section{Solution of Eq.~(\ref{eq6})}\label{AppA}
The coefficient matrix $V_{i,j}^{\alpha,\alpha'}$ and $\Gamma_{i,j}^{\alpha,\alpha'}$ in Eqs.~(\ref{eq2}) and~(\ref{eq3}) account for the real and imaginary terms of $G_{i,j}^{\alpha,\alpha'}$, respectively, given by
\begin{equation}
\begin{aligned}
V_{i,j}^{\alpha,\alpha'}=\frac{3}{4}\Gamma[\delta_{\alpha,\alpha'}(-\frac{\cos(k_{0}r)}{k_{0}r}+\frac{\sin(k_{0}r)}{k_{0}^{2}r^{2}}+\frac{\cos(ik_{0}r)}{k_{0}^{3}r^{3}})\\+\mathbf{\hat{r}}_{i,j}^{\alpha}\mathbf{\hat{r}}_{i,j}^{\alpha'}(\frac{\cos(k_{0}r)}{k_{0}r}-3\frac{\sin(k_{0}r)}{k_{0}^{2}r^{2}}-3\frac{\cos(k_{0}r)}{k_{0}^{3}r^{3}})]\\
\Gamma_{i,j}^{\alpha,\alpha'}=\frac{3}{4}\Gamma[\delta_{\alpha,\alpha'}(-\frac{\sin(k_{0}r)}{k_{0}r}-\frac{\cos(k_{0}r)}{k_{0}^{2}r^{2}}+\frac{\sin(ik_{0}r)}{k_{0}^{3}r^{3}})\\+\mathbf{\hat{r}}_{i,j}^{\alpha}\mathbf{\hat{r}}_{i,j}^{\alpha'}(\frac{\sin(k_{0}r)}{k_{0}r}+3\frac{\cos(k_{0}r)}{k_{0}^{2}r^{2}}-3\frac{\sin(k_{0}r)}{k_{0}^{3}r^{3}})]\label{eq9}
\end{aligned}
\end{equation} 

We assume the laser polarization is parallel to, for example, $\hat{z}$, the delta function $\delta_{\alpha,z}=0$ for $\alpha=\hat{x},\hat{y}$ and 1 for $\alpha=\hat{z}$. Thus, the atomic coherences $b_{i}^{\alpha}$ for $\hat{x},\hat{y},\hat{z}$ directions are written as
\begin{equation}
 \begin{aligned}
  b_{i}^{x}=\sum_{j\neq i}\frac{G_{i,j}^{x,x}b_{j}^{x}+G_{i,j}^{x,y}b_{j}^{y}+G_{i,j}^{x,z}b_{j}^{z}}{\Delta^{x}+i\Gamma/2}\\
b_{i}^{y}=\sum_{j\neq i}\frac{G_{i,j}^{y,x}b_{j}^{x}+G_{i,j}^{y,y}b_{j}^{y}+G_{i,j}^{y,z}b_{j}^{z}}{\Delta^{y}+i\Gamma/2} \\
b_{i}^{z}=\frac{\Omega^{z}e^{ik_{0}\cdot r_{i}}/2}{\Delta^{z}+i\Gamma/2}+\sum_{j\neq i}\frac{G_{i,j}^{z,x}b_{j}^{x}+G_{i,j}^{z,y}b_{j}^{y}+G_{i,j}^{z,z}b_{j}^{z}}{\Delta^{z}+i\Gamma/2}\label{eq8}
 \end{aligned}   
\end{equation}
where
\begin{equation}
 \begin{aligned}    G_{i,j}^{\alpha,\alpha'}=\frac{3}{4}\Gamma\mathbf{\hat{r}}_{i,j}^{\alpha}\mathbf{\hat{r}}_{i,j}^{\alpha'}(\frac{e^{ik_{0}r}}{k_{0}r}+3i\frac{e^{ik_{0}r}}{k_{0}^{2}r^{2}}-3\frac{e^{ik_{0}r}}{k_{0}^{3}r^{3}})
 \end{aligned}
\end{equation}
when $\alpha\neq\alpha'$ and $G_{i,j}^{\alpha,\alpha'}=G_{i,j}^{\alpha',\alpha}$. 
\begin{equation}
\begin{aligned}
G_{i,j}^{\alpha,\alpha}=\frac{3}{4}\Gamma[(-\frac{e^{ik_{0}r}}{k_{0}r}-i\frac{e^{ik_{0}r}}{k_{0}^{2}r^{2}}+\frac{e^{ik_{0}r}}{k_{0}^{3}r^{3}}\\
+\mathbf{\hat{r}}_{i,j}^{\alpha}\mathbf{\hat{r}}_{i,j}^{\alpha}(\frac{e^{ik_{0}r}}{k_{0}r}+3i\frac{e^{ik_{0}r}}{k_{0}^{2}r^{2}}-3\frac{e^{ik_{0}r}}{k_{0}^{3}r^{3}})]
\end{aligned}
\end{equation}
when $\alpha=\alpha'$. where $\mathbf{\hat{r}}_{i,j}^{\alpha}$ indicates the component of the unit vector of $\mathbf{r}_{i,j}^{\alpha}/$r along $\alpha$. The coordinate of each atom located in the optical lattice can be expressed as $(i-[i/N_{x}]N_{x}-1, [(i-N_{x}N_{y}[i/N_{x}N_{y}]-1)/N_{x}], [i-1,N_{x}N_{y}])$ in a $x-y-z$ coordinate system. For example, in a $3\times2\times2$ optical lattice shown in Fig.~\ref{fig1} (b), the coordinates for all atoms are described as (0,0,0), (1,0,0), (2,0,0), (0,1,0), (1,1,0), (2,1,0), (0,0,1), (1,0,1), (2,0,1), (0,1,1), (1,1,1), (2,1,1). Thus, the value of $\mathbf{\hat{r}}_{i,j}^{\alpha}$ can be deduced. The method of ranking atoms in the lattice is applicable throughout the model. Based on the expression of $b_{i}^{\alpha}$ in Eq.~(\ref{eq8}), the atomic coherence between any two two atoms can be solved by 3$N$-by-3$N$ linear equations. The intensity $\bf{I}(r_{s})$, therefore, can be calculated by summing up the product of $b_{i}^{\alpha}$.

\section{Fluorescence spectra of lattices}
\begin{figure}
	\centering
 \includegraphics[width=17 cm]{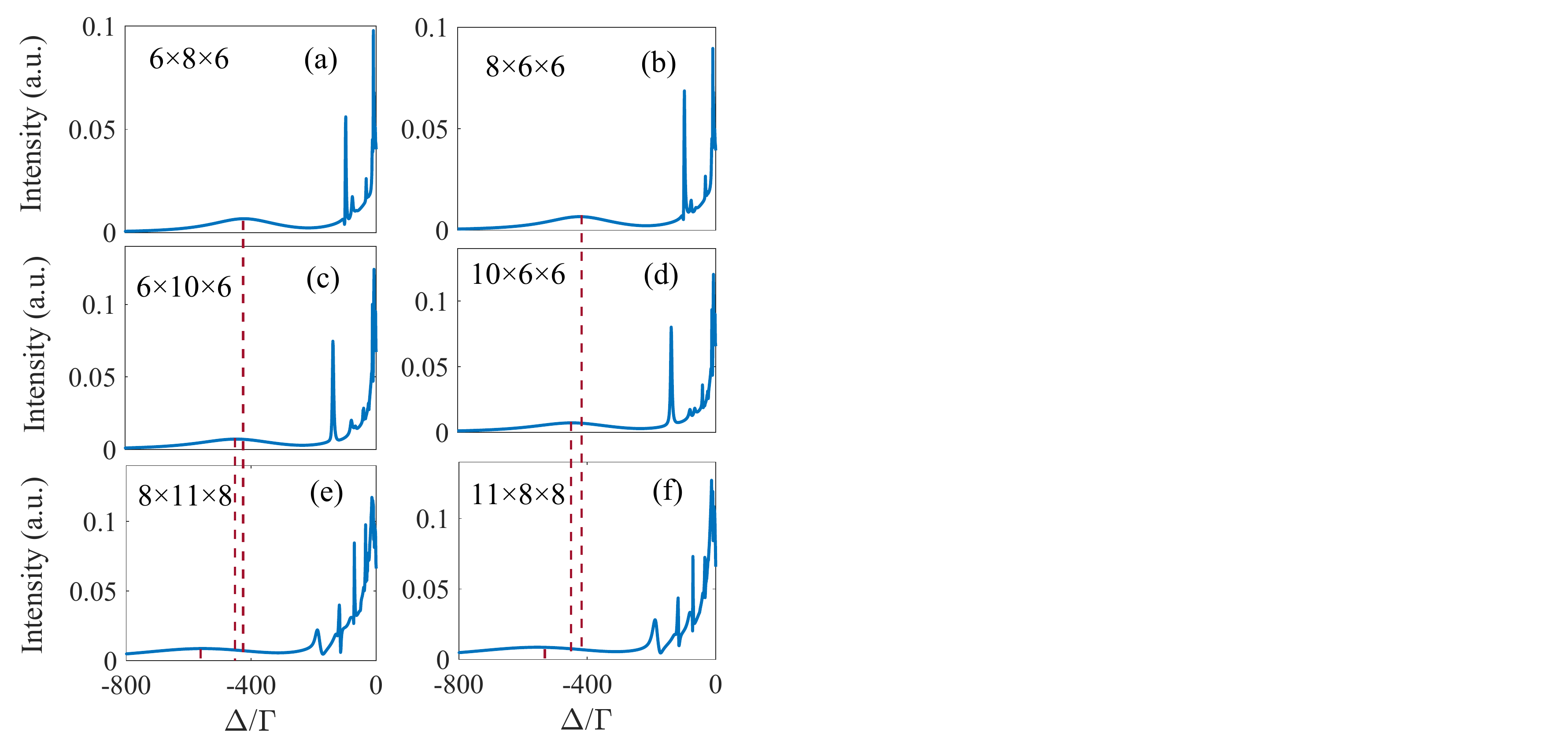}
	  \caption{Fluorescent spectra of different lattices corresponding to Fig.~\ref{fig5} at $d=a$. The narrow peaks in each sub-figure show subradiant states. The left peaks indicating superradiance exhibit features of the frequency shift and the linewidth broadening as the atom number increases. The resonance of the superradiance for each lattice is marked by a dashed line, which shows a frequency shift when atom number is enhanced. The comparison between left- and right-hand sub-figures indicates the effect of the atomic distribution in lattices.    
}\label{fig6}
\end{figure}

The scattering intensity as a function of $d$ and $\Delta$ has been studied in Fig.~\ref{fig5}. Here we fix $d=a$, and show the fluorescence spectra of the scattering for different lattices, shown in Fig.~\ref{fig6}. The frequency shift and the linewidth broadening of the spectra due to the atom-number enhancement can be observed by comparing (a), (c) and (e), or (b), (d) and (f). Moreover, the effect of the atomic distribution in lattices on the scattering intensity is revealed by the comparison of (a) and (b) or (c) and (d) or (e) and (f). Specifically, the subradiance intensity, for example, in the $11\times8\times8$ lattice is a little lower than that in the $8\times11\times8$ lattice. In addition, the subradiance is obtained in the form of sharp peaks super-positioned to the spectra. 
\end{appendices}

\medskip
\bibliography{DC}

\end{document}